\documentclass[12pt]{iopart}
\usepackage{graphicx}
\newcommand{\kbar}{\mathchar'26\mkern-9muk}
\begin{document}

\title[Diffusion Resonances in the Quantum Kicked Rotor]{Measurements of Diffusion Resonances for the Atom Optics Quantum Kicked
Rotor}
\author{M E K Williams\dag, M P Sadgrove\dag, A J Daley\dag\ddag, R N C Gray\dag, S M  Tan\dag, N Christensen\S, A S Parkins\dag and R Leonhardt\dag}

\address{\dag  Department of Physics, University of Auckland,
Private Bag 92019, Auckland, New Zealand}
\address{\ddag Institut f\"{u}r Theoretische Physik, Universit\"{a}t Innsbruck, A-6020 Innsbruck,
Austria }
\address{\S  Physics and Astronomy, Carleton College, Northfield,
MN 55057}

\begin{abstract}
We present experimental observations of diffusion resonances for
the quantum kicked rotor with weak decoherence. Cold caesium atoms
are subject to a pulsed standing wave of near-resonant light, with
spontaneous emission providing environmental coupling. The mean
energy as a function of the pulse period is determined during the
late-time diffusion regime for a constant probability of
spontaneous emission per unit time. Structure in the late-time
energy is seen to
increase with physical kicking strength. The observed structure is
related to Shepelyansky's predictions of the initial quantum
diffusion rates.
\end{abstract}

\pacs{05.45.Mt, 03.65.Yt, 32.80.Lg, 42.50.Lc}

\maketitle

\section{Introduction}
The atom-optics realization of the kicked rotor has enabled the
experimental study of the transition between quantum and classical
behaviour for this fundamental non-linear system. For example, the
effects of decoherence, the mechanism whereby quantum interference
effects are destroyed via environmental coupling \cite{Zurek},
have been studied in the quantum system. More classical-like behaviour
is observed when decoherence is added, either by spontaneous
emission events \cite{NelsonPRL,Doherty,Raizenfirstnoise},
amplitude noise \cite{Class,Class2} or timing noise \cite{Mark, NewRaizen}.
However, recent theoretical
studies have concentrated on what is perhaps a more direct
approach to studying the quantum-to-classical transition -
varying the action of the system and thereby the effective
Planck's constant, i.e., increasing the action to investigate the
limit in which `$\hbar$'$\rightarrow 0$ \cite{Kurt,AndrewD}. Of
particular interest is the behaviour found for
intermediate values of `$\hbar$', for which regions of
enhanced diffusion (known as diffusion ``resonances'')
 are predicted to exist by simulations and analytical results.

These theoretical studies have concentrated on the case
where the classical stochasticity parameter of the system stays the same
 and only the effective Planck's constant is varied.
However, in cold atom kicked rotor experiments, the most accessible parameter regime
exists for the situation where the power in the kicking
laser is held constant instead, while the pulsing period
(which is proportional to the effective Planck's constant) is changed.
 Numerically, it is found that this gives rise to different but analagous
diffusion resonances to those found in the aforementioned
simulations. Here, we present observations of
diffusion resonances in this experimentally accessible regime,
 the structure of which can be traced to a
scaling formula for the initial quantum diffusion rate derived by
Shepelyansky \cite{Shep}. The experimental
results exhibit markedly different behaviour to that predicted 
by classical calculations. An increase in the complexity of the 
resonance structure is seen as the physical kicking 
strength is increased. Experimentally, the resonances are slightly diminished
due to the non-uniform instensity profile of the kicking laser but the expected 
structure is still clearly visible.



The resonances we observe are of a different nature to
the quantum resonances previously studied in the atom optics
kicked rotor \footnote{The simplest quantum resonances
(which are also those that have been previously studied experimentally) occur
when the scaled Planck's constant is an integer multiple of $2\pi$.},
 showing non-trivial dependence on the kicking strength
as well as the scaled Planck's constant. In particular, we note that
the emphasis of our study is different from that of the related
work by d'Arcy \textit{et al.} \cite{Darcy} which focused on such quantum resonance
behaviour and not on the diffusion resonance behaviour at intermediate
values of $\hbar$. Additionally, we note the difference
between our experiments and those of  Klappauf \emph{et al.} \cite{Raizenanalomalous}
in which anomalous diffusion behaviour was studied as a function of kicking strength
for just a few values of the pulsing period. Our investigation is essentially the
converse of this experiment: Energies are measured as a function of
the pulsing period for just a few values of a kicking strength parameter,
and display a resonance structure which cannot be inferred from
previous results.

The system and model that we study is presented in Sec. II, while
the analysis of classical and quantum momentum diffusion in this
system is discussed in Sec. III. Decoherence effects due to the
small rates of spontaneous emission present in our experiments
are also considered here. Our experimental set-up is
described in Sec. IV, and the results (experimental, theoretical quantum and
classical calculations) are presented in Sec. V. The conclusion
is in Sec. VI.

\section{System and Model}

 For our system we use a laser-cooled cloud of caesium atoms of initial temperature
$\approx20$ $\mu$K interacting with a standing wave of
off-resonant laser light. The laser is pulsed with period $T$ and
pulse profile $f(t)$. If the detuning of the laser from the atomic transition
is large, the
internal atomic dynamics can be eliminated and the motion of the
caesium atoms is described by the single particle Hamiltonian \cite{Raizen}
\begin{equation}
\label{ham1} \hat{H}=\frac{\hat{p}^2}{2m}-\frac{\hbar\Omega_{\rm {eff}}}{8}
\cos(2k_{\rm{L}}\hat{x})\sum^{N}_{n=0}f(t-nT),
\end{equation}
where $\hat{x}$ and $\hat{p}$ are operators representing the
atomic position and momentum, respectively, and $k_{{\rm L}}$ is
the wave number of the laser light. The effective potential
strength,
$\Omega_{{\rm eff}}=\Omega^2(s_{45}/\delta_{45}+s_{44}/\delta_{44}+s_{43}/\delta_{43})$,
accounts for dipole transitions between different combinations of
hyperfine levels in the caesium atoms ($6{\rm{S}}_{1/2}(F=4)
\rightarrow 6{\rm P}_{3/2}(F'=5)$), where  $\delta_{ij}$ are the
corresponding detunings between laser and atomic transiton
frequencies, and $\Omega/2$ is the resonant single-beam Rabi
frequency. If we assume equal populations of atoms in all ground
state Zeeman sublevels, then $s_{45}=\frac{11}{27}$,
$s_{44}=\frac{7}{36}$, and $s_{43}=\frac{7}{108}$. It is useful to
rewrite this Hamiltonian in appropriate dimensionless units as
\begin{equation}
 \hat{\mathcal{H}}=\frac{\hat{\rho}^2}{2} - k \cos({\hat{\phi}}) \sum^{N}_{n=0}  f(\tau-n) ,
\end{equation}
which is the usual expression for the Hamiltonian of the standard
kicked rotor system. In these units - which will be referred to
as ``scaled units" - the position operator is defined by
$\hat{\phi}=2k_{{\rm L}}\hat{x}$, the momentum operator is
$\hat{\rho}=2k_{{\rm L}}T\hat{p}/m$, time is rescaled as
$\tau=t/T$, and our new Hamiltonian is related to Eq. (\ref{ham1})
by $\hat{\mathcal{H}}=(4k_{{\rm L}}^2T^2/m)\hat{H}$. The
classical stochasticity parameter is given by
$\kappa=\Omega_{{\rm eff}}\omega_{{\rm r}}T\tau_p$, where
$\tau_{\rm p}$ is the pulse length in unscaled time and $\omega_{{\rm r}}=\hbar
k_{{\rm L}}^2/2m$. In our experiments, $f(\tau)$ is a good
approximation to a square pulse, i.e., $f(\tau)=1$ for
$0<\tau<\alpha$, where $\alpha=\tau_{\rm p}/T$. Note that
$k=\kappa/\alpha$.

In scaled units we have $ [\hat{\phi},\hat{\rho}]=i\kbar$, with
$\kbar=8\omega_{{\rm r}}T$, so that the quantum behaviour of our
system is reflected by an effective Planck's constant, $\kbar$ ,
which increases as we increase the pulse period $T$. This
reflects our ability to change the total action in the system,
and hence how classically our system behaves (for larger $\kbar$
values the quantum nature of the system should be more apparent).
Note that the effective Planck's constant is proportional to the
ratio of the total classical action of the system to $\hbar$.

The natural experimental unit for momentum is that of two photon
recoils,  $2\hbar k_{{\rm L}}$, and $p/(2\hbar k_{{\rm L}})$
will henceforth be referred to as the momentum in experimental
units. We note the relationship $\rho/\kbar=p/(2\hbar
k_{{\rm L}})$ and also define the quantity
$\phi_{{\rm d}}=\kappa/\kbar=\Omega_{{\rm eff}} \tau_p/8$ as a
dimensionless measure of the physical kicking strength.
Experimentally, it is easier to hold this quantity constant,
rather than $\kappa$, as $T$ is varied, because a constant value of
$\phi_{{\rm d}}$ corresponds to constant pulse duration, standing
wave detuning and power (whereas $\kappa$ is proportional to $T$).

Our system is coupled to its environment via atomic spontaneous
emission events, which occur when the caesium atoms absorb photons
from the standing wave \cite{NelsonPRL} and then spontaneously
re-emit the photons in random directions. We characterise the
level of this decoherence by the probability of spontaneous
emission per atom per kick, $\eta$. Given the large detuning used
in our experiments,
i.e., $\Omega_{{\rm eff}}/\delta \ll 1$, this process may be
modeled by the following master equation for the density operator
$\hat{w}$ of the system \cite{Doherty},
\begin{eqnarray}
\label{master1}
\dot{\hat{w}}&=&-\frac{{\rm i}}{\kbar} [\hat{\mathcal{H}},\hat{w}]-\frac{\eta}{\alpha}\sum^{N}_{n=0} f(\tau-n)[\cos^2(\hat{\phi}/2),\hat{w}]_+\nonumber \\
& &+ 2\frac{\eta}{\alpha}\sum^N_{n=0}f(\tau-n)\int^1_{-1}{\rm d}u N(u){\rm e}
^{{\rm i} u \hat{\phi}/2} \nonumber \\
& &\times \cos(\hat{\phi}/2)\hat{w}\cos(\hat{\phi}/2){\rm e}^{-{\rm i} u
\hat{\phi}/2},
\end{eqnarray}
where $N(u)$ is the distribution of recoil momenta projected onto
the axis of the standing wave, and $[.,.]_+$ denotes an
anti-commutator. Simulations of Eq. (\ref{master1}) are used for
comparisons with the experiment.

\section{Momentum Diffusion}

We measure the total kinetic energy of the cloud after $N$ kicks,
which depends on the initial energy of the cloud plus the
increase in the kinetic energy resulting from the kicks. The
amount of increase for kick number $n$ is the momentum diffusion
rate, given by
$2D(n)=\langle\hat{\rho}_{n+1}^2\rangle-\langle\hat{\rho}_n^2\rangle$,
where we denote $\hat{\rho}_0=\hat{\rho}(t'=0)$,
$\hat{\rho}_1=\hat{\rho}(t'=1)$, etc.. For a kicked rotor system
with a sufficiently broad initial momentum distribution, we
expect $D(0)=D(1)=\kappa^2/4$. The system then passes through an
initial quantum diffusion period lasting typically for around 5
kicks \cite{AndrewD}, with a diffusion rate approximated by the
result of Shepelyansky (under the conditions $\kbar \geq 1$ and
$\kappa \gg \kbar$) \cite{Shep},
\begin{equation}
\label{shepform}
D_{\rm q}\!=\!\frac{\kappa^2}{2}\!\left(\!\frac{1}{2}\!-\!J_2(K_{\rm q})\!-\!J_1^2(K_{\rm q})\!+\!J_2^2(K_{\rm q})\!+\!J_3^2(K_{\rm q})\!\right),
\end{equation}
where $K_{\rm q}=2\kappa \sin(\kbar/2)/ \kbar$. Note that the
long time classical diffusion rate is also given by Eq. (\ref{shepform}),
but with $K_{\rm q}\rightarrow\kappa$ (i.e.,
$\kbar\rightarrow0$).

Without decoherence, the system generally settles into a
localised state \cite{Cohen}, but the loss of phase coherence
produced by the addition of spontaneous emission causes the
system to settle instead into a final steady state diffusion
regime, with a late time diffusion rate which may be approximated
by the formula \cite{NelsonPRL,AndrewD,Cohen}
\begin{equation}
\label{sumldr} D_{\infty}=\sum^{\infty}_{n=0}\eta(1-\eta)^nD_0(n),
\end{equation}
where $D_0(n)$ is the diffusion rate at the $n$th kick for a
kicked rotor \textit{without} decoherence. Essentially, this
formula assumes that dynamical correlations over particular time
intervals which give rise to the late time diffusion rates are
suppressed by a factor equal to the probability that a
spontaneous emission occurs within that time interval. The
correlations taken over a set number of kicks give rise to the
diffusion rates seen in the kicked rotor without decoherence
after that number of kicks, which leads to the late time
diffusion rate being an appropriate weighted average over the
diffusion rates as the kicked rotor ``settles down"
\cite{AndrewD}. Thus, the diffusion rates in the first few kicks
are essentially ``locked in" by the spontaneous emission events,
and we observe similar structure in the late time diffusion rates
as we vary $T$ to that observed in the initial quantum diffusion
rates.

\begin{figure}
\centering
\includegraphics[height=10cm]{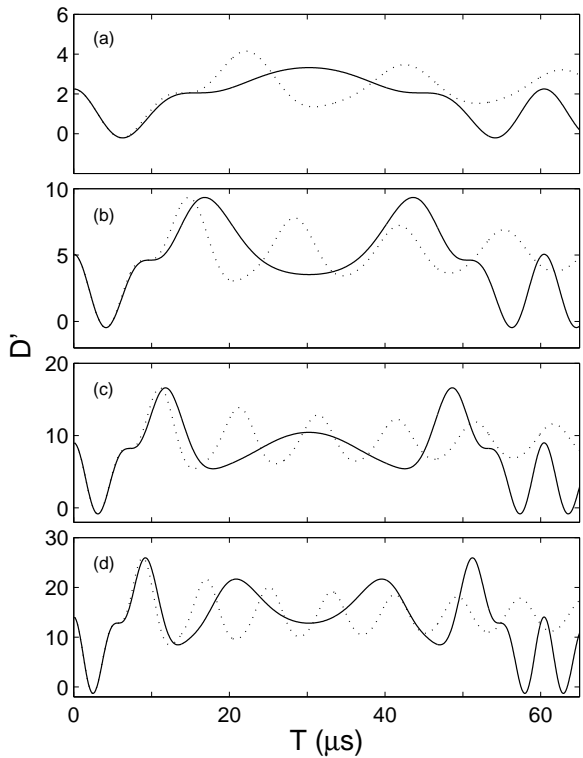}

\caption{Theoretical initial momentum diffusion rates in the
quantum case (solid line), $D_{\rm q}'$ as given by Eq.
(\ref{Dq}), and the classical case (dotted line), $D_{\rm cl}'$
in Eq. (\ref{Dcl}). The rates are given in experimental units as a function of $T$
for (a) $\phi_{{\rm d}}=3$, (b) $\phi_{{\rm d}}=4.5$,
 (c) $\phi_{{\rm d}}=6$ and (d) $\phi_{{\rm d}}=7.5.$ The quantum
 resonance is seen in the quantum diffusion rate at $T=60.4$ $\mu$s.} \label{fig:shepphid}

\end{figure}

The structure we observe in the initial quantum diffusion rates
as we vary $T$ for constant $\phi_{{\rm d}}$ with diffusion
rates measured in experimental units is particularly interesting.
We can express Shepelyansky's formula in this regime as
\begin{equation}
\label{Dq}
D_{\rm q}'\!=\!\frac{(\phi_d)^2}{2}\!\left(\!\frac{1}{2}\!-\!J_2(K_{\rm q}')\!-\!J_1^2(K_{\rm q}')\!+\!J_2^2(K_{\rm q}')\!+\!J_3^2(K_{\rm q}')\!\right),
\end{equation}
with $K_{\rm q}'=2 \phi_{{\rm d}} \sin(4\omega_{{\rm r}}T)$.
We then see that any structure in the diffusion rates is periodic
in $T$ with period $2 \pi/8\omega_{{\rm r}}$ (In fact, from Eq.
(\ref{sumldr}) this is also true for the late time diffusion
rates). We also see that the form of the structure depends solely
on the value of $\phi_{{\rm d}}$. Fig. \ref{fig:shepphid} shows
the initial quantum diffusion rate as a function of pulse period
for varying values of $\phi_{{\rm d}}$. We see the regular
feature of a peak at the quantum resonance when
$\kbar=2\pi$ ($T=60.4$ $\mu$s), and we see
increasing numbers of pronounced peaks or diffusion resonances as we
increase the value of $\phi_{{\rm d}}$.

The classical diffusion rate can be similarly found in this
regime to be
\begin{equation}
\label{Dcl}
D_{\rm cl}'\!=\!\frac{(\phi_d)^2}{2}\!\left(\!\frac{1}{2}\!-\!J_2(\kappa)\!-\!J_1^2(\kappa)\!+\!J_2^2(\kappa)\!+\!J_3^2(\kappa)\!\right),
\end{equation}
with $\kappa=8\omega_{{\rm r}}T\phi_{{\rm d}}$.  The dotted
curves in Fig. \ref{fig:shepphid} show the classical rates for
various values of $\phi_{{\rm d}}$. These rates oscillate around the
quasilinear value which in these units is $(\phi_{{\rm d}})^2/4$,
with the oscillations increasing in frequency with increasing
$\phi_{{\rm d}}$. For any given $\phi_{{\rm d}}$ however, the
structure in the classical diffusion rate is markedly different to
that in the initial quantum diffusion rate.

In the experiment, late-time energies were measured and not
initial diffusion rates. However, the diffusion resonances
discussed above are still observable. The existence of the same
structures in the late time diffusion rates (in the presence of
decoherence) as those in the initial quantum rates has been
verified using the simulations described later in this paper. An
approximate formula for the energy after $N$ kicks can be found
by summing the diffusion rate at each kick, that is
\begin{equation}
\label{Energy}
E'(N)=\frac{\langle\left(\frac{p(N)}{2\hbar
k_{{\rm L}}}\right)^2\rangle}{2}=\sum^{N-1}_{n=0}D'(n,T).
\end{equation}
Therefore, with the initial and final diffusion rates both
displaying these structures, it follows that the energy at the
$N$th kick should also display them.

\section{Experimental Setup}
\label{Sec:ExpSetup}
The experimental setup used was much the same as that used
previously in our quantum chaos investigations \cite{Nelson,
NelsonPRL}, with a few modifications. A standard six-beam
magneto-optical trap (MOT) was used to trap and cool approximately
$10^{5}$ caesium atoms. The trapping laser frequency was set
about $10$ MHz to the red of the $6{\rm S}_{1/2}
(F=4)\rightarrow6{\rm P}_{3/2}(F'=5)$ transition. A second
(repump) laser was locked to the
$6{\rm S}_{1/2}(F=3)\rightarrow6{\rm P}_{3/2} (F'=4)$
transition to return those atoms lost to the $F=3$ ground state
to the trapping cycle. After a $20$ ms cooling phase prior to
kicking, the cloud had a temperature of approximately $20$ $\mu$K
and a width of $\sigma_{{\rm cloud}}\sim270$ $\mu$m in its
position distribution. Kicking of the cloud occurred for up to
$2$ ms during the $10$ ms free expansion phase, at the completion
of which the cloud was ``frozen" in space by the molasses beams
and imaged. The repumping beam was left on during the kicking to
prevent loss of atoms to the $F=3$ ground state. The resultant
heating effect was negligible for our experiments.

A third laser was used to create a pulsed optical standing wave
across the cloud. A relatively high power laser diode was
injection locked with a frequency stabilised external cavity
laser, giving a beam of up to $22$ mW CW power. For fast
switching the beam passed through a $80$ MHz Acousto-Optic
Modulator (AOM) in front of a single mode polarisation preserving
optical fibre. Temporal modulation was provided via the RF supply
to the AOM, generating pulse shapes very close to rectangular. The
linearly polarised beam was then collimated resulting in a beam radius
at the cloud of $2\sigma_{{\rm beam}}=1$ mm. Finally, to create a
standing wave the beam was retroreflected by a mirror outside the
vacuum cell. The atoms experienced a range of optical potential
depths as the cloud's width was comparable in size to that of the
laser beam. If $\phi_{{\rm d,max}}$ is the kicking strength along
the beam axis then the mean value was found to be
$\phi_{{\rm d,mean}} \approx\ 0.77\phi_{{\rm d,max}}$ with a
standard deviation of $18\%$. This effect and additional factors
leading to a slightly greater spread in $\phi_{rm d}$ values is discussed below.
 In the following, $\phi_{{\rm d}}$
will always refer to $\phi_{{\rm d,mean}}$. The detuning of the
kicking beam to the blue of the $F=4 \rightarrow F'=5$ transition
was monitored by overlapping the trapping and kicking beams and
monitoring the resultant beat signal. Both the beam detuning and intensity
were chosen to give a desired $\phi_{{\rm d}}$ while maintaining
a constant spontaneous emission rate. The range of
$\phi_{{\rm d}}$ examined in this way was from $\phi_{{\rm d}}
=3.3$ to $6.6$. Taking reflection losses at the cell windows into
account, over this $\phi_{{\rm d}}$ range the Rabi frequency
varied from $\Omega/2\pi=34-76$ MHz with corresponding detunings
of $\delta _{45}/2\pi= 315-740$ MHz, thus $\delta>>\Gamma,\Omega$
(where $\Gamma$ is the linewidth of the atomic transition)
 was satisfied for all $\phi_{{\rm d}}$ values considered.

For a chosen $\phi_{{\rm d}}$ the pulse length was held constant
($\tau_{{\rm p}}=520$ ns), while the pulse period was varied
from $2.5$ $\mu$s to just above the quantum resonance at
$T\approx60$ $\mu$s. Thirty kicks were delivered to the cloud for
each pulse period. The images of the expanded cloud were averaged over
the dimension perpendicular to the kicking beam to yield
momentum distributions for the kicked cloud and
the mean energy $E=\langle p^2\rangle/2$ was
calculated for each distribution. High momenta have a large effect on these energy
values and as this was where the signal dropped for the higher
energy kicked clouds, much care was taken to reduce the effects
of noise. For an experimental run involving a single
$\phi_{{\rm d}}$ value, the subtracted background was an average
from just before and just after the run. Any slight fluctuations in
background level were accounted for by defining the zero level
for each momentum distribution via an image taken just before
commencement of the kicking sequence, omitting the small cloud.
The signal-to-noise ratio was on average about 100:1 and for each
value of $T$ the momentum distribution was measured 5 times. To minimise the
effects of long term fluctuations in the kicking beam, one fibre end was angle cleaved and the
beam power was checked and readjusted several times throughout a
run. These measures reduced the fluctuations in $\phi_{{\rm d}}$
from this source to $\sim1\%$.

\begin{figure}
\centering
\includegraphics[height=7.5cm]{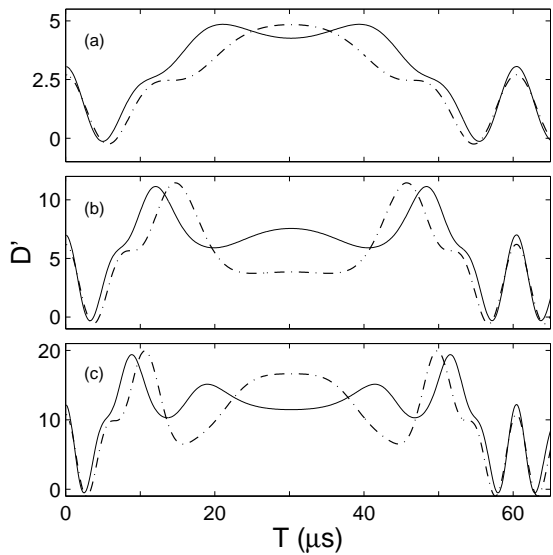}
\caption{Initial momentum diffusion rates, $D'$, for constant
$\phi_{{\rm d}}$ (dash-dotted line) and for
$\phi_{{\rm d}}$ with a distribution centered on $\phi_{{\rm d, max}}
=\phi_{{\rm d}}/0.77 $ with standard deviation $0.18\phi_{\rm{d}}$ (solid line).
$D'$ is given in experimental units as a function
of $T$ for (a) $\phi_{{\rm d}}=3.3$, (b) $\phi_{{\rm d}}=5.0$ and
(c) $\phi_{{\rm d}}=6.6$.} \label{fig:whyprob}
\end{figure}

We now examine the cause of the spread in kicking strengths.
Firstly, the finite width of the kicking beam results in atoms at
different radial positions across the beam interacting with laser
fields of different intensity.
The effect of the spread in kicking
strengths on the initial Shepelyansky diffusion rates is easily
examined and also applies to the late-time energies. If $\rho(r)$
is the 2-dimensional cloud density as a function of the radial coordinate $r$,
and $\phi_{{\rm d}}(r)$ the distribution of kicking strengths,
then we can approximate the diffusion rate over all of the particles
and for a given
$\phi_{{\rm d},{\rm max}}$ as \cite{Mark}
$
\label{diffvary}
\bar{D}(\phi_{{\rm d},{\rm max}},T)=\int_{0}^{\infty}D(\phi_{{\rm d}}(r),T)\rho(r)2\pi
r{\rm d}r,
$
where $D(\phi_{{\rm d}}(r),T)$ is given by Eq. (\ref{Dq}).
Calculating $\bar{D}(\phi_{{\rm d},{\rm max}},T)$ for a broad
kicking beam with $\sigma_{{\rm cloud}}<<\sigma_{{\rm beam}}$,
corresponding to constant $\phi_{{\rm d}}(r)$, reproduces the
pronounced structure as seen in Fig. \ref{fig:shepphid} and in the
simulations. But calculating the diffusion rate with the 2:1
beam-to-cloud width ratio as used experimentally and with
$\phi_{{\rm d},{\rm max}}=\phi_{{\rm d}}/0.77$, gives less
accentuated diffusion resonance structure which is broader over
the pulse period values. Fig. \ref{fig:whyprob} displays these
results for a few values of $\phi_{{\rm d}}$. Thus, a spread in
$\phi_{{\rm d}}$ values creates an averaging effect which
somewhat diminishes the height of the resonance structure
and shifts the positions of the resonant peaks relative
to the case for a single $\phi_{\rm{d}}$ value.

Secondly, a spread in $\phi_{{\rm d}}$ values is also caused by atoms in
different magnetic substates of the $F=4$ level coupling to the
higher energy states with different transition strengths,
resulting in atoms in different substates experiencing different
kicking strengths.

The combination of this effect with that caused
by the finite beam width can create a spread of $\phi_{{\rm d}}$
values with a standard deviation as large as $20\%$ of the mean $\phi_{{\rm d}}$
value. In order to account for this, we performed additional
simulations in which the $\phi_{{\rm d}}$ value for each
trajectory was chosen from a distribution based on the 2:1
beam-to-cloud width ratio for our system (assuming that each has a
radial Gaussian profile). Also included were the relative
coupling strengths of atoms in different magnetic substates for
the experimental detunings $\delta=315,385,485,575,740$ MHz for
$\phi_{{\rm d}}=3.3, 4.0, 5.0, 5.9, 6.6$ respectively (assuming an
equal population of atoms in each substate).

\section{Experimental Results}

\begin{figure*}
\centering
\includegraphics[width=16cm]{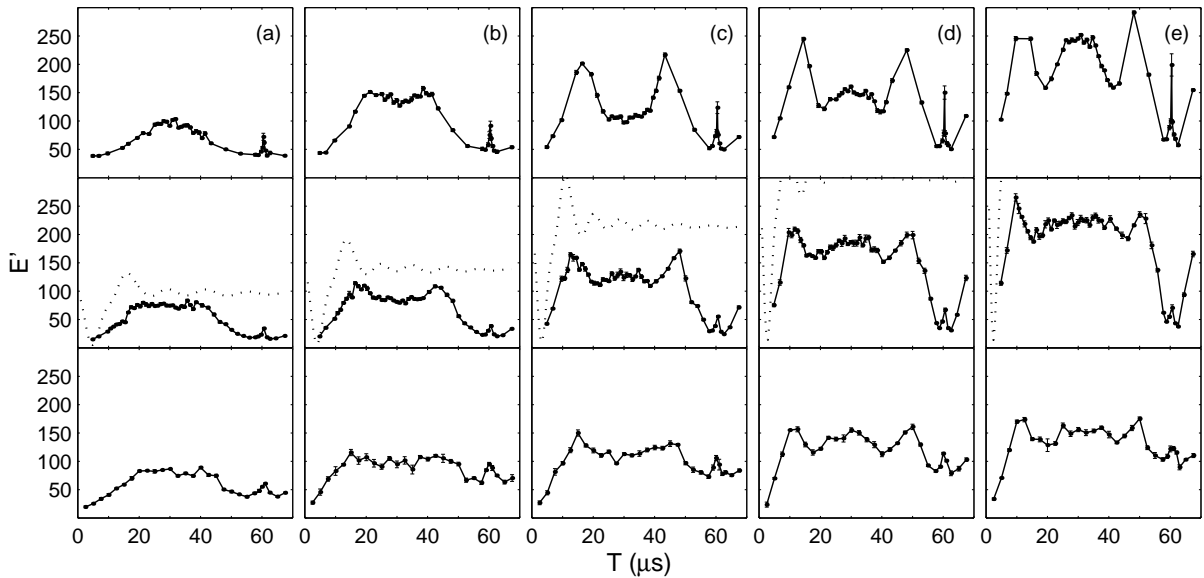}
\caption{Quantum simulations with fixed $\phi_{{\rm d}}$ (top) and
the corresponding experimental results (bottom) giving the energy,
$E'$, after 30 kicks as a function of pulse period with
$\tau_{{\rm p}}=520$ ns, $\eta=0.0125$ for (a) $\phi_{{\rm
d}}=3.3$, (b) $\phi_{{\rm d}}=4.0$, (c) $\phi_{{\rm d}}=5.0$, (d)
$\phi_{{\rm d}}=5.9$ and (e) $\phi_{{\rm d}}=6.6$. Additional
quantum simulations (middle, solid line) take into account the
spread in physical kicking strength $\phi_{{\rm d}}$, as do the
analytical classical results (middle, dotted line). The energies
are in experimental units and error bars for the experimental
and simulation results are shown but are very small.} \label{fig:expres}
\end{figure*}

Fig. \ref{fig:expres} shows results from numerical simulations as
well as those from experimental measurements. The first row shows
simulated energies for different, well defined values of $\phi_{\rm d}$.
All simulations were performed using the
 Monte Carlo wavefunction method as in Ref. \cite{AndrewD}. The
simulations are for a system with an initial Gaussian
momentum distribution of width
$\sigma_{\rho}/\kbar=\sigma_p/2\hbar k_l=4$ (this corresponds to
a cloud of temperature $\approx 20 \mu K$).

To take into account the unavoidable spread of $\phi_{\rm d}$ values in
our experiments, simulations were also performed in which the kicking
strength for each trajectory was sampled from the theoretical distribution
of $\phi_{\rm d}$ experienced by a spherical cloud of atoms
(see Section \ref{Sec:ExpSetup}), and the energy
for each value of $T$ was taken to be the incoherent average of the energies
over all such realisations. The results of these simulations are seen in the
second row of Fig. \ref{fig:expres}. Comparison of the first two rows shows that,
while the non-uniformity of the kicking beam intensity makes the diffusion
resonances somewhat less distinct, the structure of interest is still clearly
visible and thus amenable to experimental investigation.

Comparison of experimental results
with those from the simulations (bottom and middle rows of
Fig. \ref{fig:expres} respectively) shows very good agreement between the two.
The quantum resonance at
$T \approx 60$ $\mu$s is seen to be present
for all $\phi_{{\rm d}}$ values. Additionally, while a single broad
peak similar to that seen by d'Arcy \textit{et al}. \cite{Darcy}
is found for $\phi_{{\rm d}}=3.3$, for larger $\phi_{{\rm d}}$
values more complicated structure is observed. This peak splits
into two peaks which then diverge from each other, whereupon a third
peak rises between them. The structure
mirrors that in the initial quantum diffusion rate as given by
Shepelyansky's result in Eq. (\ref{Dq}) and shown in Fig.
\ref{fig:shepphid}.


There is still some discrepancy in energy values between
the experimental results and the additional simulations. For
$\phi_{{\rm d}}=5.9$ and $6.6$ the measured energies are, overall,
lower than expected, while for all $\phi_{{\rm d}}$ they
are larger than in the simulations around the quantum
resonance region of $T=58$-$68$ $\mu$s. The first problem is
accounted for by realising that in recording the momentum
distributions, at some point signals at higher momenta fall below
the noise level of the CCD. Hence, the measured total energy of the cloud is
systematically lower than the true energy after kicking. For
larger $\phi_{{\rm d}}$, for which energies are higher in
general, this problem is particularly pronounced as more atoms
lie in the wings of the distribution.

The discrepancy around the
quantum resonance currently remains unexplained, but could be a
systematic effect related to the larger pulse period values in
this region. Continuing investigations will hopefully resolve this
issue. However, we would like to emphasise that the energy structure
seen at values of $T$ lower than that at which quantum resonance occurs
is of primary interest to us (experimentally, the quantum resonance
itself provides a convenient
confirmation of the pulse timing calibration).

Overall, we find very good qualitative agreement between experimental results
and our simulations. The expected structure is evident
and its complexity increases with $\phi_{{\rm d}}$ as
in simulations. In particular, the resonant peaks are seen where
they are expected for each value of $\phi_{{\rm d}}$.
For comparison, the energies of the classical system
after 30 kicks were also computed by assuming that the diffusion
rate for the first two kicks is $\phi_{{\rm d}}/4$, and for
subsequent kicks is given by Eq. (\ref{Dcl}). They were averaged
over the same spread in $\phi_{{\rm d}}$ values that was used in
the additional quantum simulations and are shown in the middle row
of Fig. \ref{fig:expres}. Note that the energies are larger than
the quantum simulations, as would be expected, and for
$\phi_{{\rm d}}=5.9$ and $6.6$ go off the scale, oscillating
around $E'=300$ and $560$ respectively.

The classical energies
clearly exhibit very different behaviour from those measured in
the quantum system. The difference becomes more marked
at higher values of $T$ (i.e. when the system behaves more quantum mechanically)
where the oscillations in the classical energy die down, whilst
pronounced diffusion resonances and quantum resonances are seen in the quantum system.
  This can be contrasted with the results of Klappauf
\emph{et al.} where, in both the classical and quantum
systems, the energy oscillates about the quasilinear value and only a
relative shift in the peak positions demarcates the two
regimes. The striking difference between the quantum behaviour and the
expected classical behaviour in these experiments makes them a thorough
testing ground for the effects of decoherence on the quantum kicked rotor. Indeed,
preliminary investigations into the effects of noise on these
diffusion resonances \cite{AuckUnpub} suggest that quantum and diffusion
resonances are affected differently by noise induced decoherence for a
quantum kicked rotor with environmental coupling.

\section{Conclusion}

We have presented experimental and simulation
results showing non-trivial behaviour in the late-time energy (and
thus diffusion rate) of the atom optics kicked rotor for values
of the pulse period intermediate between those periods associated with
quantum resonances. The observed
resonances are distinct from the quantum resonances previously studied
in the kicked rotor, being dependent on the kicking strength as well
as the value of the pulse period.

The non-trivial features predicted by quantum simulations are matched
very closely by our experimental measurements.
 The relationship between the observed
structure and Shepelyansky's scaling for the quantum correlations
is evident. Furthermore, we note that the structure observable in
the late time energies confirms that in a system subject to
environmental coupling a dependence of the late-time diffusion
rate on the initial quantum diffusion rate exists.

In addition, the observed
energy measurements show a very different trend to that
predicted by classical calculations. With the introduction of noise,
these experiments should provide a novel testing ground
for the efficacy of decoherence in regaining the classical limit, and
we have begun preliminary studies in this area.

We thank Maarten Hoogerland for interesting discussions related to our experimental investigations.
This work was supported by the Royal Society of New Zealand Marsden Fund, grant UOA016.


\end{document}